\documentclass[preprint,,showpacs,showkeys,aps,endfloats]{revtex4}
\usepackage{graphicx}
\usepackage{amssymb}

\def\ahe3{$^3$He}
\def\he4{$^4$He}
\def\o16{$^{16}$O}
\def\al27{$^{27}$Al}
\def\xhe3ag{$^3He(\alpha,\gamma)^7Be$}
\def\s34{S$_{34}$}

\begin{document}

\title{In Beam Tests of Implanted Helium Targets}
\thanks{Work Supported by USDOE Grant Nos: DE-FG02-94ER40870, DE-FG02-91ER40609, DE-FG02-97ER41033, and DE-FG02-97ER41046.}

\author{J.E. McDonald$^{1}$, R.H. France III$^{2}$, R.A. Jarvis$^3$, M.W. Ahmed$^4$, M.A. Blackston$^4$, Th. Delbar$^5$, M. Gai$^{3,6}$, T.J. Kading$^3$, Y. Parpottas$^4$, B.A. Perdue$^4$, R.M. Prior$^7$, D.F. Rubin$^6$, M.C. Spraker$^7$, J.D. Yeomans$^1$, L. Weissman$^3$, H.R. Weller$^4$, and E.L. Wilds Jr$^3$.}

\affiliation{1. Department of Physics, University of Hartford, 
200 Bloomfield Ave, West Hartford, CT 06117-1599, USA.}
\affiliation{2. Department of Chemistry and Physics, Georgia College and State University, 
Campus Box 82, Milledgeville, GA 31061-0490, USA.}
\affiliation{3. Laboratory for Nuclear Science at Avery Point, University of Connecticut, 
1084 Shennecossett Rd, Groton, CT 06340-6097, USA.}
\affiliation{4. Department of Physics, Duke University, and Triangle Universities Nuclear Laboratory,Box 90308, Durham, NC 27708-0308, USA.}
\affiliation{5. Institut de Physique Nucleaire, Universite Catholique de Louvain, 2 chemin du cyclotron, B-1348 Louvain-la-Neuve, Belgium.}
\affiliation{6.  Department of Physics, Yale University, PO Box 208124, 
272 Whitney Avenue, New Haven, CT 06520-8124, USA.}
\affiliation{7. Department of Physics, North Georgia College and State University, Dahlonega, Georgia, 30597, USA.}

\begin{abstract}

Targets consisting of $^{3,4}$He implanted into thin aluminum foils (approximately 100, 200 or 600 $\mu$g/cm$^2$) were prepared using intense (a few $\mu$A ) helium beams at low energy (approximately 20, 40 or 100 keV). Uniformity of the implantation was achieved by a beam raster across a 12 mm diameter tantalum collimator at the rates of 0.1 Hz in the vertical direction and 1 Hz in the horizontal direction. Helium implantation into the very thin (approximately 80-100 $\mu$g/cm$^2$) aluminum foils failed to produce useful targets (with only approximately 10\% of the helium retained) due to an under estimation of the range by the code SRIM. The range of low energy helium in aluminum predicted by Northcliffe and Shilling and the NIST online tabulation are observed on the other hand to over estimate the range of low energy helium ions in aluminum. An attempt to increase the amount of helium by implanting a second deeper layer was also carried out, but it did not significantly increase the helium content beyond the blistering limit ($~6 \ \times \ 10^{17}$ atoms/cm$^2$). The implanted targets were bombarded with moderately intense \he4 and $^{16}$O beams of 50-100 particle nA . Rutherford Back Scattering of 1.0 and 2.5 MeV proton beams and  recoil helium from 15.0 MeV oxygen beams were used to study the helium content and profile before, during and after bombardments. We observed the helium content and profile to be very stable even after a prolonged bombardment (up to two days) with moderately intense beams of $^{16}$O or \he4. Helium implanted into thin (aluminum) foils is a good choice for thin helium targets needed, for example, for a measurement of the \xhe3ag reaction and the associated $S_{34}$ astrophysical cross section factor (S-factor).
 \end{abstract}
 \pacs{25.55.-e, 26.65.+t, 28.41.Rc, 29.25.Pj}
\keywords{implanted target, helium target, fusion of \ahe3 + \he4, Rutherford Back Scattering}

\date{\today}
\maketitle

\section{Introduction}

Several key reactions of interest in nuclear physics and nuclear astrophysics involve light particles such as protons, deuterons, and alpha particles. The use of radioactive beams and inverse kinematics makes the availability of targets containing these elements necessary. Previous experiments in the field have used a variety of cryogenic, solid, and gas targets. One solution that has been used for hydrogen and deuterium is a thin polyethylene target. This solution cannot be used for helium, however, and some published experiments have used gas cells for helium targets.  A helium gas jet target that was used in a measurement of the \xhe3ag reaction \cite{Kre82} was found later to have significant non-uniformity and yield a cross section which is approximately 40\% smaller than measured by the same group in a later experiment \cite{Hil88}. Another method that has shown promise is the implantation of helium into a thin metal foil ({\em e.g.}, Al, Ni, Cu, Mo, Au). This has been used in a number of experiments, such as measurements of the \xhe3ag reaction cross section \cite{Al84}, as well as the $^3He(d, p)^4He$ reaction cross section \cite{Ge96}. Noble gases, such as Ar and Xe \cite{Sp93,Gr95} implanted into metals were also used as targets. Indeed, noble gasses are known to form stable structures composed of bubbles with radii of approximately 100 nm when implanted into metals.

An implanted target  suffers from a low number density, approximately a factor of 10 below targets commonly used in reaction studies in nuclear astrophysics, but has many advantages over a gas cell, with the most significant advantage being a well-localized target with high energy resolution. For example, a \ahe3 implanted target with $5 \times 10^{17} \  atoms/cm^2$ (just below the blistering limit) and a 1.0 MeV \he4 beam of modest intensity of 500 nA (280 pnA) yield a count rate of 40 counts per hour in a HPGe detector with a 2\% efficiency for detecting the resulting 2.016 MeV direct capture gamma rays at 90$^\circ$. The large angular acceptance of such a detector placed close to the target results an energy spread with a FWHM of approximately 30 keV. Hence an implanted target in combination with moderately intense beams is useful for studying the \xhe3ag reaction and its associated astrophysical cross section factor ($S_{34}$) at modestly low energies of approximately $E_{cm}$ = 400 keV. 

One area of concern with implanted targets is the stability of the implanted gas over time, especially when used in beams. Beam heating has been cited as a problem \cite{Al84}. For example it was noted \cite{Al84} that the helium content in the center of the target was depleted after bombardment with an intense (2-3 $\mu$A) \he4 beam even though the implanted target was water cooled. 

In this study we used helium implanted into thin aluminum foils (approximately 100, 200 or 600 $\mu g/cm^2$) in order to minimize the beam heating. We tested these thin implanted targets before and after bombardment with moderately intense beams (approximately 100 pnA ) of $^{16}$O and \he4. We did not observe a depletion in the helium content even after a prolonged (as much as two day) bombardment. The helium content and profile measured using back scattering of 1.0 MeV proton beams were observed to be very stable.

\section{Target Preparation}

The targets were prepared both at the Institute of Material Science at the University of Connecticut (UConn) using a Varian CF-3000 Ion Implanter and at the 80 KV low energy (polarized ion source) facility of the Triangle University Nuclear Lab (TUNL) at Duke University.  The implanter used at UConn employed a commercial dose processor and uniformity monitor to evenly distribute a precise amount of the desired ion in the target material. The TUNL implantation was performed using a beam raster in both the vertical (0.1 Hz) and horizontal (1 Hz) direction. The $^4$He implanted at UConn was into an Al foil  $600 \ \mu g/cm^2$ thick, which corresponds to $1.33 \times 10^{19} \ atoms/cm^2$, and $^{3,4}He$ atoms were implanted in the TUNL facility into approximately 80, 100, and 210 $\mu g/cm^2$ aluminum foils. The thin aluminum foils (approximately 100 $\mu g/cm^2$) were prepared by evaporation and the 216 $\mu g/cm^2$ foils were purchased from Good Fellow Corp. They were mounted on frames with a 25 mm diameter open area and implanted area with a 12 mm diameter.

The range and straggle of the implanted ions were estimated using the code SRIM \cite{SRIM}. For the UConn \he4 implantation, an energy of 100 keV was chosen (which also corresponds to the lowest energy available from that implanter). At 100 keV the helium range is predicted by SRIM to be 153.6 $\mu g/cm^2$ (5688 A). Four implantations were performed at the TUNL laboratory: 20 keV  \he4 beams were implanted into 82, 94, and 120 $\mu g/cm^2$ aluminum foils; 30 keV \ahe3 beams were implanted into 112 and 119 $\mu g/cm^2$ aluminum foils; 25 keV \ahe3 beams were implanted into 100 and 102$\mu g/cm^2$ aluminum foils; and 45 keV \ahe3 beams were implanted into 216 $\mu g/cm^2$ (0.0008 cm) thick aluminum foils. A majority of the results discussed in this paper were obtained using targets with 216 $\mu g/cm^2$ aluminum foils. All implantations used a fluence of approximately $4 \times 10^{17} \ ions/cm^2$-- just below the blistering limit of $6 \times 10^{17} \ He/cm^2$ indicated in the literature \cite{Fo76,Wa79,Da79a,Da79b,Fe80,Ca83}. The implantation fluence was measured by integrating the implanted beam. However problems with the suppression of secondary electrons prohibited us from measuring the total fluence with high accuracy.

In the implantation performed at UConn, we confirmed that the implanted helium did not remain in the aluminum foil when the areal density exceeded the blistering limit. In an attempt to increase the helium content in the foils, we prepared targets with two layers of $^4$He.  This was done by two distinct methods.  In the first method, a layer was implanted using an implantation energy of 160 keV followed by a second layer at 100 keV.  In the second method, a layer was implanted at 100 keV and then the target was flipped so that the second layer at 160 keV was implanted from the opposite side.  Although the areal density for each layer was below the blistering limit, the total for both layers exceeded the blistering limit value of  $6 \times 10^{17} \ ions/cm^2$.  Spectra from a test of these two-layer targets showed low statistics and very broad $^4$He peaks, indicating a low distributed helium ion content in the foils.  This was attributed to the probability that the layers did not remain distinct, and together exceeded the blistering limit for $^4$He in Al. During the TUNL implantation, we also produced one target with half of the dose implanted from one side of the foil and the second dose from the back side. The results obtained with this target are discussed below.

\section{Target Tests and Results}

Several tests of the implanted targets were performed using Rutherford Back Scattering (RBS) of 1.0 MeV proton beams extracted from the 1 MV single ended Yale Teaching Accelerator Lab, 2.5 MeV proton beams extracted from the TUNL tandem accelerator, and low energy oxygen beams extracted from the Yale ESTU tandem. The measured yield for scattering from the target material is given by:

\hspace{1 in} $Y(\Theta) \ = \ N_{beam} \ \times \ N_{target} \ \times \ {d\sigma \over d \Omega _{cm}} (\Theta ) \times \ {d \Omega _{cm} \over d \Omega _{Lab}}(\Theta ) \ \times d \Omega _{Lab}$
\    \\
\   \\
Note that neglecting to consider the Jacobian = ${d \Omega _{cm} \over d \Omega _{Lab}}(\Theta )$ for p + He scattering (of the order of 0.5) led to the wrong conclusion that only 50\% of the implanted helium dose is retained in the target \cite{We00}. The cross section ${d\sigma \over d \Omega _{cm}}(\Theta )$ is either the calculated RBS cross section when appropriate, or the measured cross section for p + \al27 \cite{Ch01,Ra02}, p + \ahe3 \cite{Me68}, or the cross section calculated from the measured phase shifts of the elastic scattering of p + \he4 \cite{Br67}.
 
The test at the ESTU tandem at the A.W. Wright Nuclear Structure Laboratory at Yale University used a 15 MeV $^{16}$O beam.  The elastic scattering of $^{16}$O from the aluminum substrate was measured using detectors placed at $\Theta_{Lab} \ = \ 30 ^\circ, \   and \  = \ 40 ^\circ$ and the recoil $^4$He were detected using silicon surface barrier detectors placed at $\Theta_{Lab} \ = \ 10 ^\circ, \ 15 ^\circ, \ 20 ^\circ, \ 25 ^\circ, \ 40 ^\circ, \ and \ 45 ^\circ; \ \Theta_{cm} \ = \ 180^\circ - 2\Theta_{Lab}$, with an aluminum absorber foil ($2.7 \ \mu g/cm^2$) which stopped the scattered oxygen ions and aided in the separation of  alpha particles from protons.  The recoil alpha particles lost about 1 MeV in the absorber foil, contributing about 100 keV to the FWHM energy resolution of the recoil alpha spectra.  The detectors were also collimated to an area of 100 $mm^2$, causing them to subtend an angle of $\pm 1 ^\circ$, contributing 270 keV to the FWHM energy resolution. We note that the measured FWHM of about 520 keV (see Fig.1) is dominated by the energy loss of the $^{16}$O beam traversing the implanted layer of approximately 360 keV FWHM. 

The measured helium recoil spectra are shown in Fig. 1.  The extracted target content shown in Fig. 2 was found to be independent of scattering angle, confirming the assumption of Rutherford scattering of \he4 + $^{16}O$.  The areal density of $^4$He was measured to be $3.4 \pm 0.3 \times 10^{17} \ atoms/cm^2$, which is equal to the implanted fluence. We do not confirm the suggested loss of helium discussed in Ref. \cite{We00}.  The aluminum areal density was measured using RBS of the oxygen beam to be $1.40 \pm 0.1 \times 10^{19} \ atoms/ cm^2$, which confirmed the manufacturer's quoted foil thickness of $600 \ \mu g/cm^2$.

The helium profile was measured by measuring the FWHM of the spectra of recoil helium from a 20 MeV $^{16}$O beam with an average intensity of 50 particle nA. As shown in Fig. 3 the FWHM remained constant over the measurement duration of 18.4 hours indicating a stable helium profile during this prolonged bombardment with an oxygen beam of moderate intensity.  In Fig. 4 we show the ratio of the yield of recoil $^4$He to elastic scattering of $^{16}$O from aluminum plotted against the cumulative oxygen ions beam dose. No deterioration of the target was observed for a continuous exposure to beam up to 18.4 hours at this beam intensity.

All \ahe3 and \he4 implanted targets were tested prior to bombardment by the scattering of a very low intensity 2.5 MeV proton beam (approximately 30 nA) into two detectors placed at +173$^\circ$ and -173$^\circ$ with respect to the beam. The spectra obtained are shown in Fig. 5. One of the targets was implanted with 50\% of the dose ($2 \ \times \ 10^{17} \ helium/cm^2$) from the front side and the same dose from the back of the target. The low energy loss of 2.5 MeV protons in aluminum did not allow separation of protons scattered off the two separated helium layers, as shown in Fig. 5. But the larger energy loss of 1.0 MeV protons allowed us to observe two distinct peaks as shown in Fig. 6. The doubly implanted target was exposed over two days to \he4 beams having an average intensity of 80 particle nA . This two layered and all the single layered targets (that were not exposed to beams) were tested by scattering 1.0 MeV proton beams into a detector placed at 163$^\circ$. The obtained spectra are shown in Fig. 6.

We first note that the SRIM calculations \cite{SRIM} predict a range of 92.8 $\mu g/cm^2$ (3436 A) for 50 keV \he4 in \al27. But the Northcliffe and Schilling (NS) tabulation \cite{North} predicts a range of 142 $\mu g/cm^2$ and the NIST \cite{NIST} online tabulation predicts a range of 141.8 $\mu g/cm^2$. Our measurements indicate that the SRIM calculations underestimate the range of very low energy alpha particles in aluminum and the NS and NIST tabulations overestimate the range of low energy alpha-particles in aluminum.

As an example the SRIM calculations predict a range of 44.4 $\mu g/cm^2$ (1645 A) for 20 keV \he4 in aluminum, but the NIST tabulation predicts a range of 83.7 $\mu g/cm^2$ at this energy. The NS tabulations does not list a range for energies below 50 keV. For the TUNL implantation of 20 keV \he4 into the very thin foils (80 and 100 $\mu g/cm^2$) the SRIM calculations place the implanted \he4 in the middle of the aluminum foil, but the NIST estimate places it at the end of these foils. The test of these very thin implanted targets using back scattering of a 1.0 MeV proton beam revealed that only 10\% of the \he4 fluence was retained in the aluminum foil, suggesting an underestimate of the range by SRIM. For the 45 keV \ahe3 beam (15 keV/amu) SRIM predicts a range of 106.3 $\mu g/cm^2$ (3937 A) with a straggle of 27.4 $\mu g/cm^2$ (1013 A). The SRIM calculations predicts that the two sided implantation of 45 keV \ahe3 into the 216 $\mu g/cm^2$ aluminum foil would yield two helium layers right on top of each other in the center of the doubly implanted target. The NIST tabulation predicts the larger range of 156.5 $\mu g/cm^2$ thus two helium layers separated by 110 $\mu g/cm^2$. In Fig. 6 we observe two distinct peaks from the scattering of 1.0 MeV which is at odd with the SRIM prediction. These two low energy peaks were fitted  resulting two Gaussians separated by 40 keV, corresponding to two \ahe3 layers separated by 45 $\mu g/cm^2$ (note that at backward angles the energy loss is doubled), which is smaller than the separation predicted by the NS and the NIST tabulations. The FWHM of the second narrow peak (Fig. 6b) and the narrow peak of the single layer target (Fig. 6a) of 27 keV correspond to 31 $\mu g/cm^2$ of \al27 and is, in fact, consistent with SRIM prediction for the straggle of helium in aluminum.

It is also worth noting that the first peak at low energy is broader (46 keV) than the second one (27 keV). Since the first peak corresponds to the first implantation from the front, we conclude that the second implantation caused the broadening of the first implanted layer. While the first peak is broadened, the layer still contains the same amount of \ahe3 as the first layer, see Fig. 7. This observation allows us to conclude that the implantation beam of 45 keV and intensity of 2-3 $\mu$A with a beam heating of approximately 100 mW, does not remove the helium content but only causes a broadening of the profile. This observation is similar to the effect observed with the heating of a $^7Be$ implanted target recently reported by the Weizmann group \cite{Weiz}. The 1.0 MeV \he4 beams that one may use for measuring the \xhe3ag reaction with a 216 $\mu g/cm^2$ \ahe3 implanted target lose 375 keV in the target (placed at 45 $^\circ$ with respect to the beam). Hence our observations suggest that beam heating will not be a problem even for a beam with a modest intensity of 265 particle nA (approximately 500 nA). As shown above such a beam target combination yields a useful count rate for measuring the \xhe3ag reaction at energies as low as approximately $E_{cm} \ = \ 400$ keV.

The \ahe3 implanted into 216 $\mu g/cm^2$ foils were scanned with 1.0 MeV proton beams with a diameter of 3 mm using eight steps of 2.4 mm each. This scan across the doubly-implanted target was performed after two days of bombardment with 80 particle nA beam of \he4 with a diameter of approximately 6 mm in the center of the 12 mm implanted helium area. The  results of the scan are shown in Fig. 7. We note that the ratio of the yield of scattered protons from \ahe3 and \al27 does not vary across the doubly implanted target, nor does the peak position or the FWHM of the two observed peaks. This suggest the helium profile and content of the two layers remained stable even after the two days of bombardment.

From the known cross section for scattering of p + \ahe3 \cite{Me68}, the Rutherford scattering of p + \al27 \cite{Ch01,Ra02}, the Jacobian, and simple kinematics we calculate the yield ratio for \al27/\ahe3 to be 3.99 times the atomic number ratio. The \ahe3/\al27 yield ratio of 1.73\% leads to \al27/\ahe3  number ratio of 14.5 or a \ahe3/\al27 number density of 6.9\%. With an aluminum thickness of 216 $\mu g/cm^2$ and number density of $4.8 \ \times \ 10^{18} \ /cm^2$ we derive a helium double layer including $3.3 \ \times \ 10^{17} \ ^3He/cm^2$, consistent with the implanted fluence (the integrated beam).

\section{Conclusion}

We prepared $^{3,4}$He targets by implanting low energy helium beams into thin aluminum foils and tested them using (mostly Rutherford) Back-Scattering of low energy proton and oxygen beams before, during and after bombardment with modestly intense beams of \he4 and $^{16}$O.  We measured the helium profile and content Using (mostly Rutherford) Back-Scattering and demonstrated that these targets are robust against bombardment by modestly intense beams. No  degradation in the helium profile or content was observed after a prolonged bombardment with moderately intense beams. Such targets can be very useful for certain studies with modestly intense beams, including a measurement of the \xhe3ag reaction at energies as low as $E_{cm}$ = 400 keV.

\section{Acknowledgment}

The author thank the accelerator staff of WNSL at Yale, TUNL at Duke, and Lawrence Cerrito of the Yale 1 MV Teaching Accelerator Lab for the delivered beams.

\begin{figure}
\includegraphics[width=4in]{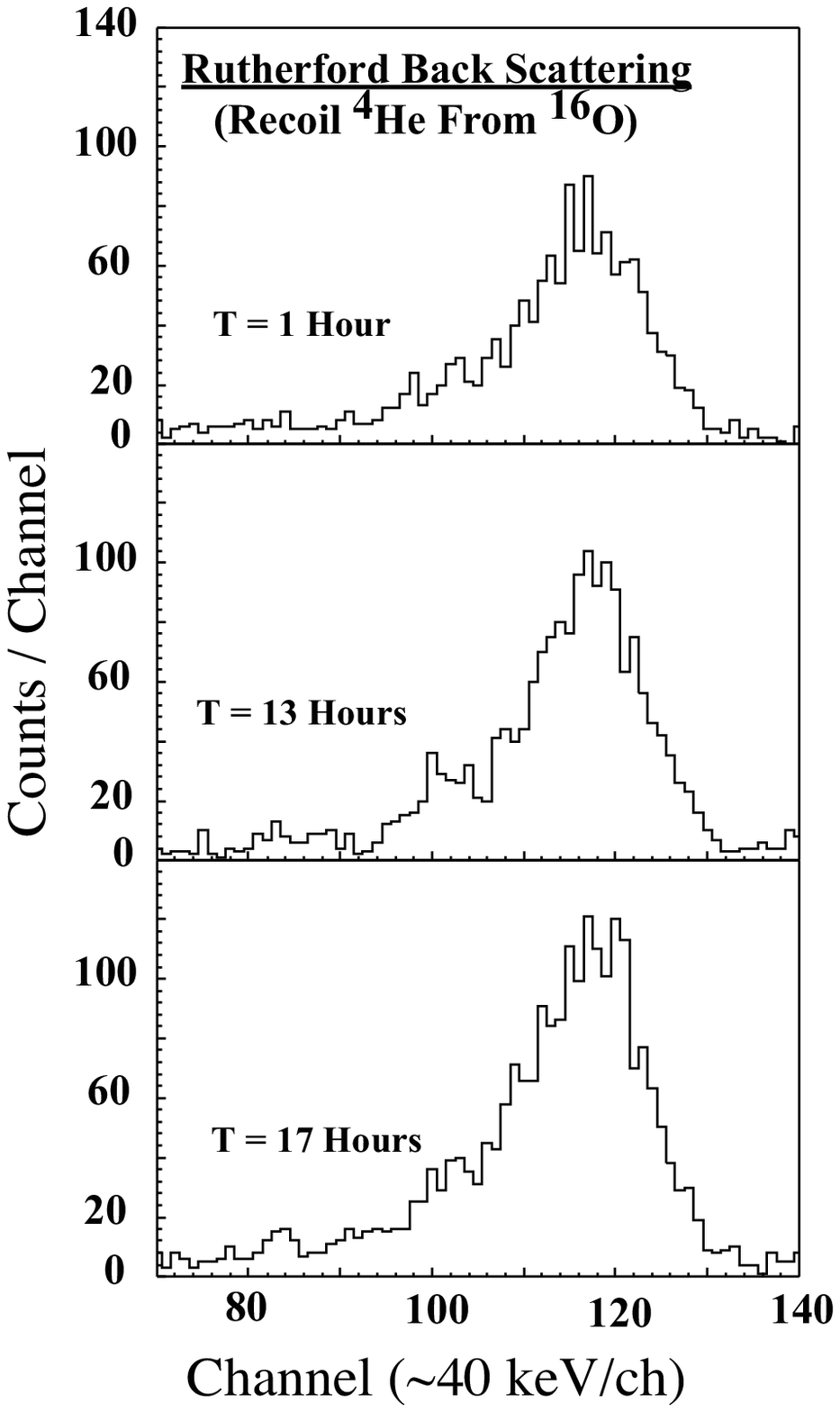}
 \caption{\label{Recoil} The measured spectra of recoil helium from 15.0 MeV oxygen beams. Spectra taken with similar integrated oxygen beams at 1, 13 and 17 hours are shown.}
\end{figure} 

\begin{figure}
\includegraphics[width=7in]{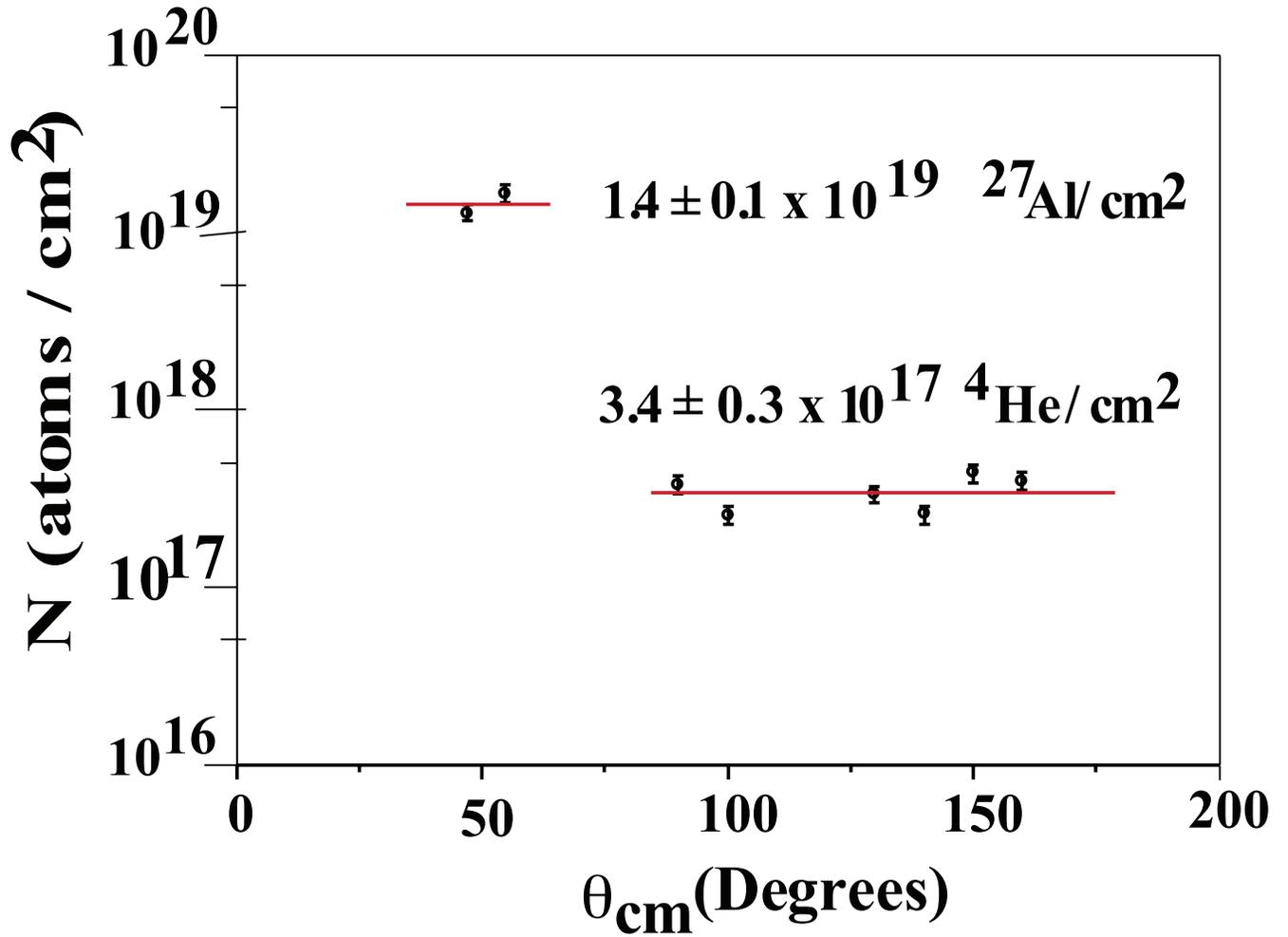}
 \caption{\label{Content} The \he4 content of the implanted target prepared at UConn and measured at Yale with low energy oxygen beams.}
\end{figure}

\begin{figure}
\includegraphics[width=7in]{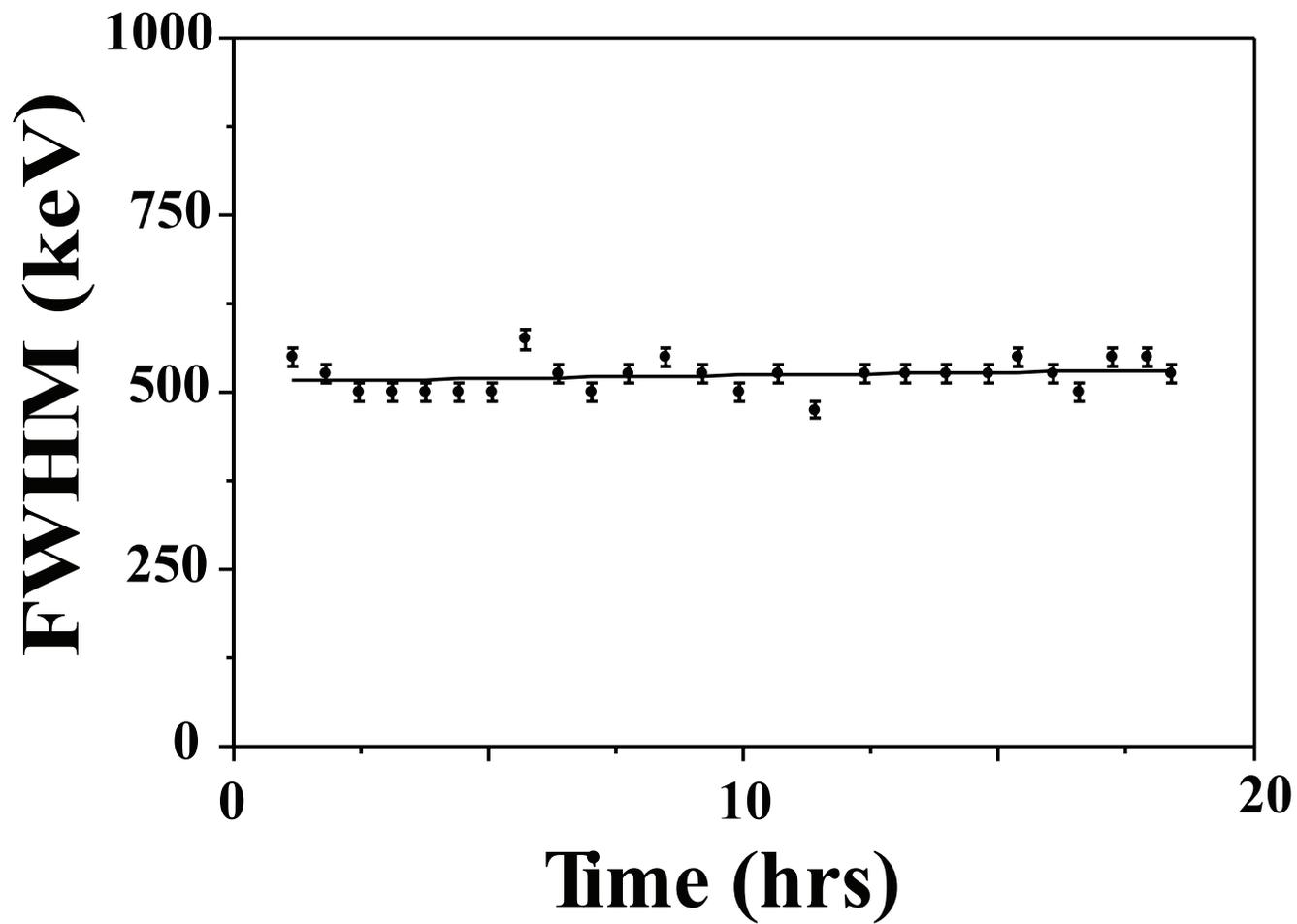}
 \caption{\label{FWHM1} The measured FWHM of the recoil \he4 shown in Fig. 1 as a function of time.}
\end{figure}

\begin{figure}
\includegraphics[width=7in]{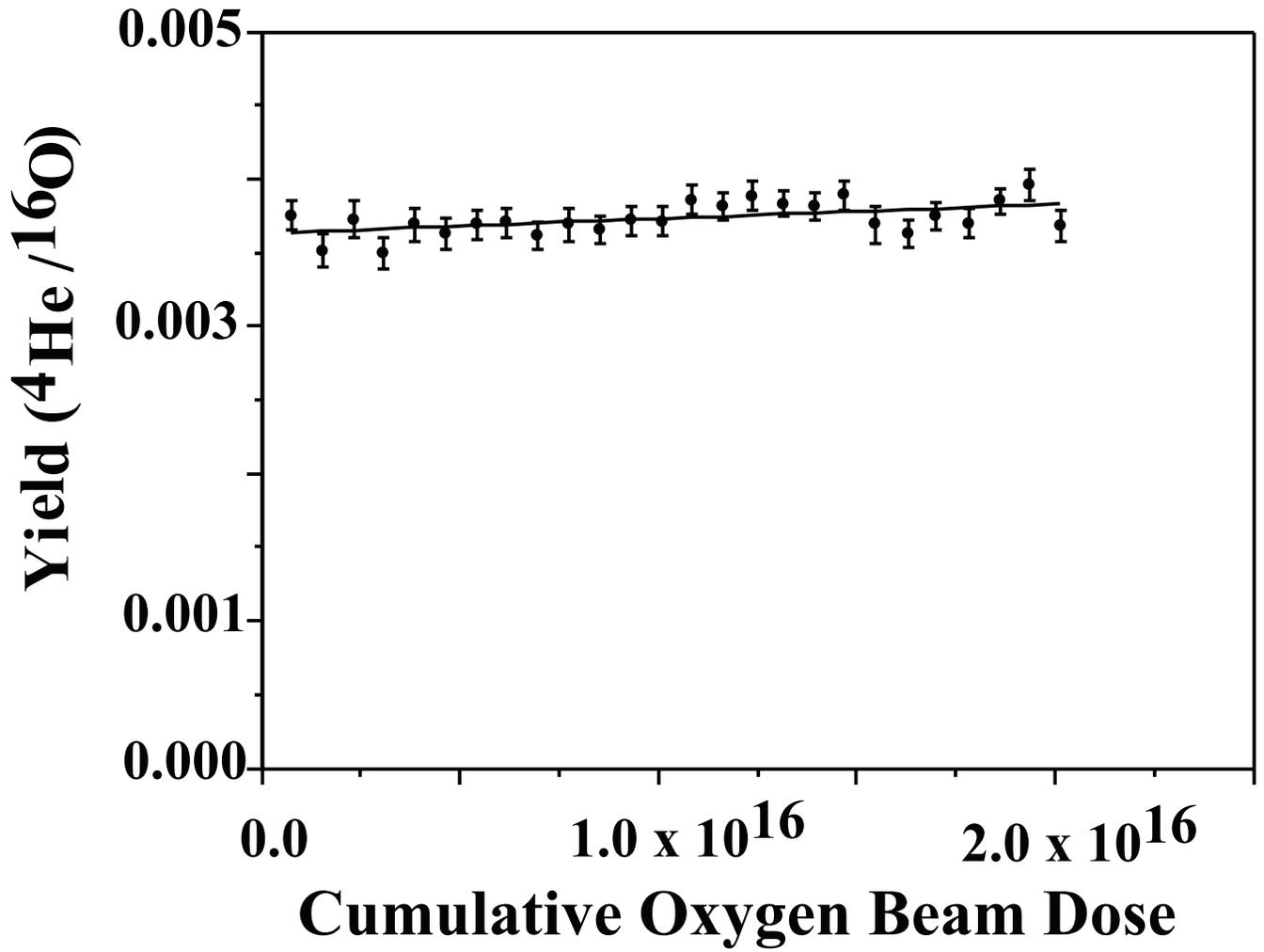}
 \caption{\label{Ratio} The ratio of the measured yield of recoil \he4 and elastic scattering of oxygen from \al27 as a function of cumulative oxygen beam dose.}
\end{figure}

\begin{figure}
\includegraphics[width=6in]{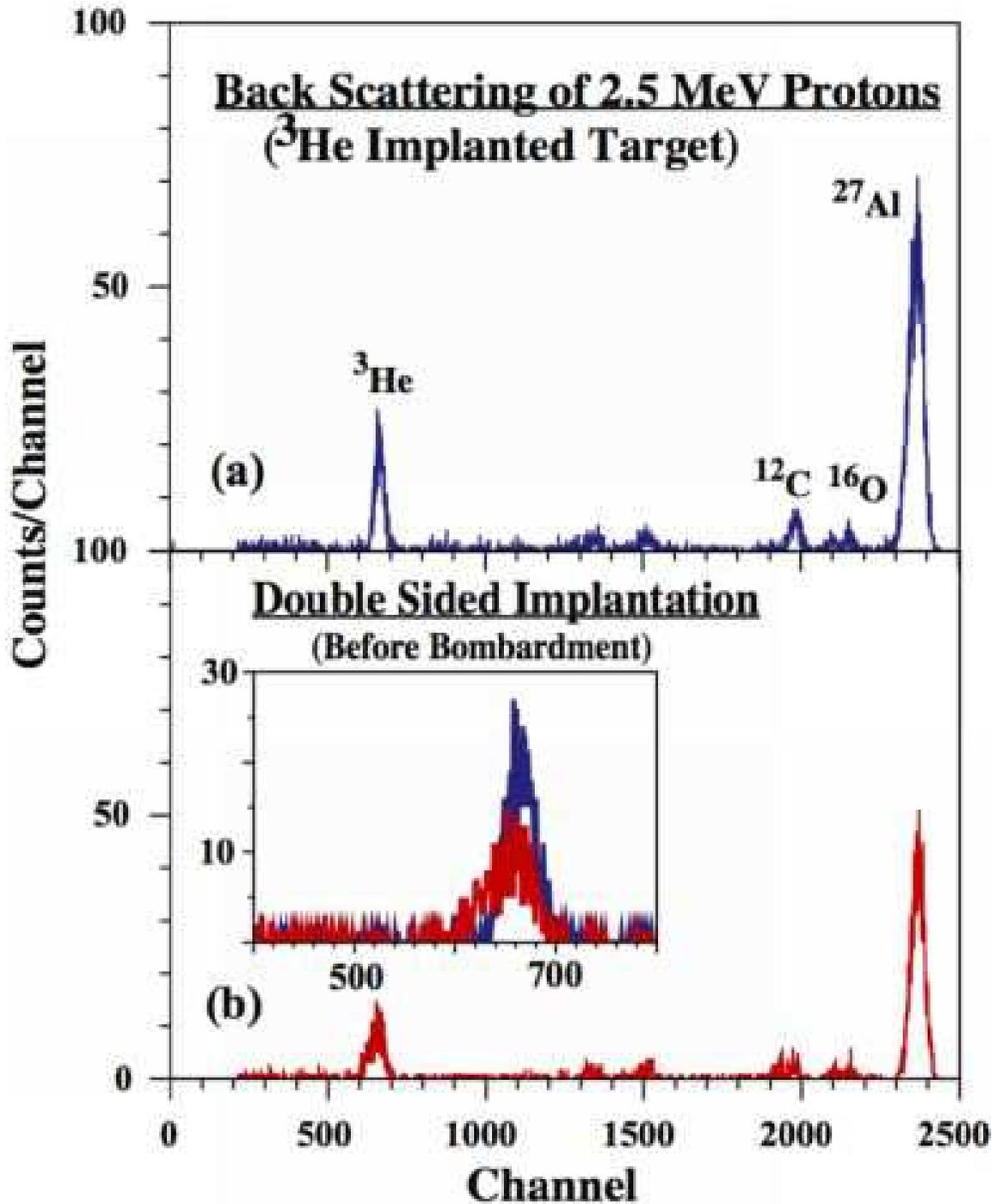}
 \caption{\label{2MeV} The measured spectra of elastic scattering of 2.5 MeV proton beams from implanted targets prepared at TUNL: (a) single implanted $^3$He target before bombardment (blue color) and (b) double implanted $^3$He target (red color), also before bombardment.}
\end{figure}

\begin{figure}
\includegraphics[width=6in]{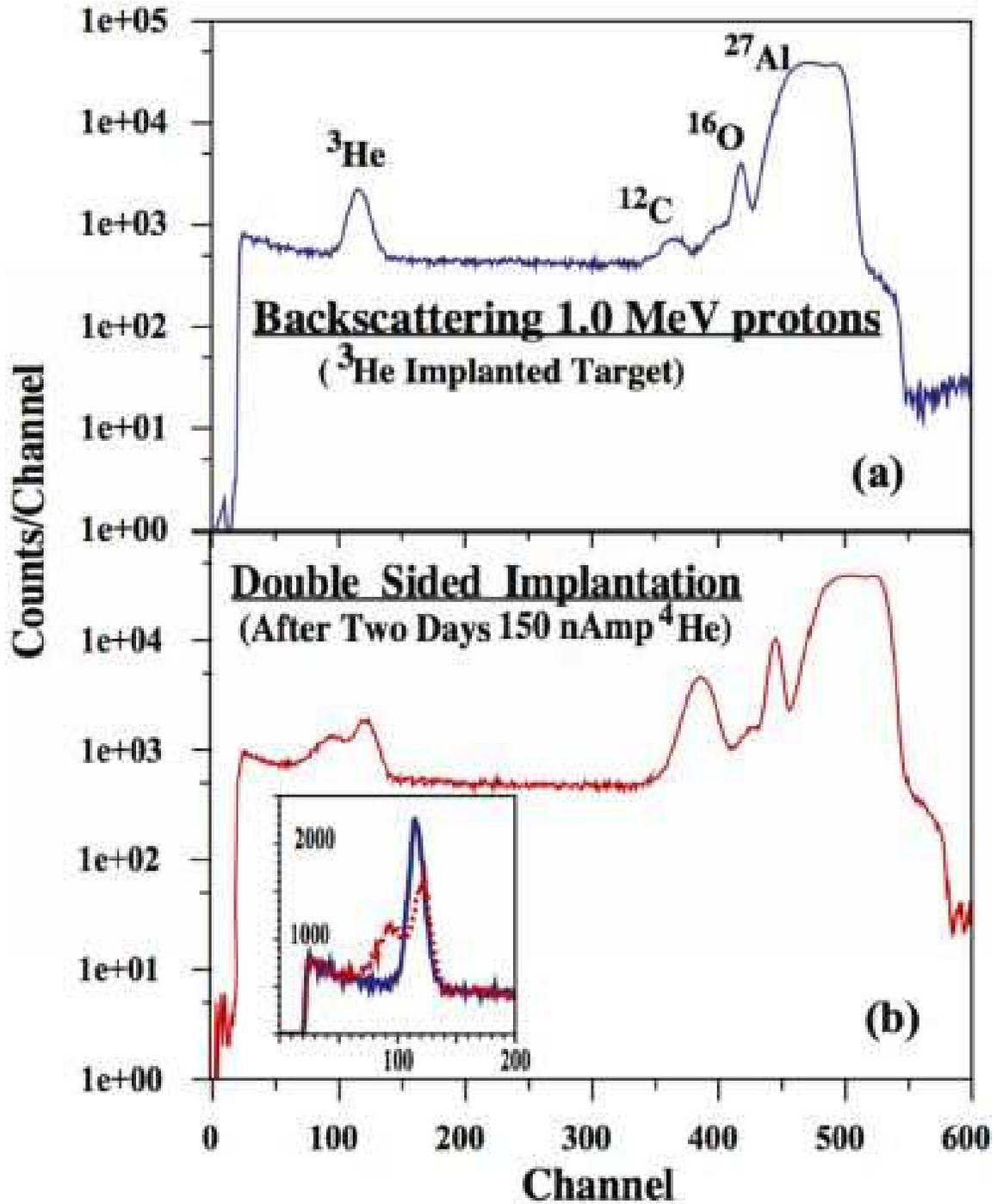}
 \caption{\label{1MeV} The measured spectra of elastic scattering of 1.0 MeV proton beams from (a) single implanted $^3$He target before bombardment (blue color) and (b) double implanted $^3$He target after bombardment (red color). The observed carbon buildup in spectrum (b), is approximately 5 $\mu g/cm^2$ after two days of bombardment.}
\end{figure}

\begin{figure}
\includegraphics[width=4.5in]{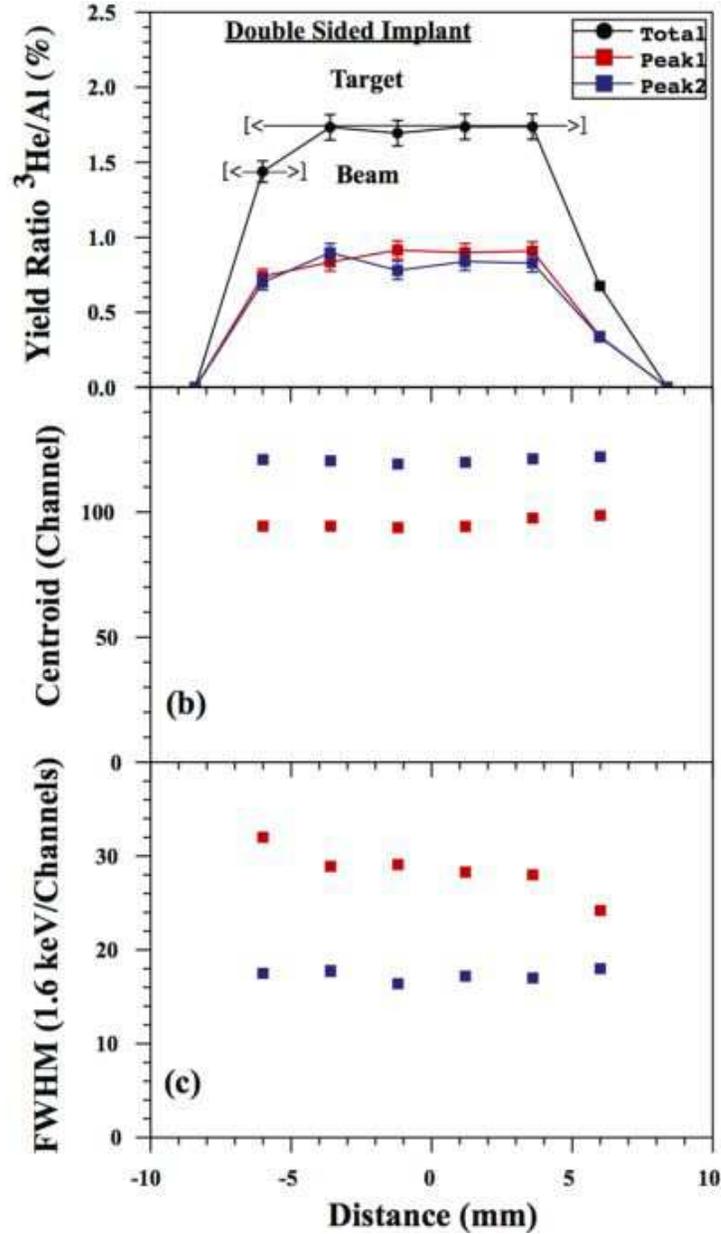}
 \caption{\label{Scan} Results of the scan across the double sided $^3$He implanted target using 1.0 MeV proton beams. We show (a) the ratio of yield of elastic scattering from $^3$He and \al27 for the two low energy peaks shown in Fig 6b, corresponding to the two helium layers and the total sum of yield, (b) the centroid of the two peaks, (c) the FWHM of the observed peaks. The scan was performed after a two day bombardment with \he4 beams as discussed in the text. The $^3$He implanted area ($\pm$6 mm from target center), and the 1.0 MeV proton beam spot (diameter of 3 mm) are shown. Lines are drawn to connect data points and guide the eye.}
\end{figure}

\end{document}